\documentclass[published]{JHEP3} 

\JHEP{00(2011)000}

\JHEPspecialurl{http://jhep.sissa.it/JOURNAL/JHEP3.tar.gz}

\usepackage{epsfig,multicol,bbm}
\usepackage{amssymb}

\newcommand\fverb{\setbox\fverbbox=\hbox\bgroup\verb}
\newcommand\fverbdo{\egroup\medskip\noindent%
			\fbox{\unhbox\fverbbox}\ }
\newcommand\fverbit{\egroup\item[\fbox{\unhbox\fverbbox}]}
\newbox\fverbbox

\usepackage[latin1]{inputenc}

\newcommand{\LL}{\mathcal{L}}

\title{ {\bf Palatini Actions and Quantum Gravity Phenomenology}}

\author{Gonzalo J. Olmo \\ 
	Departamento de Física Teórica and IFIC, Centro Mixto Universidad de Valencia - CSIC.
  Facultad de Física, Universidad de Valencia, Burjassot-46100, Valencia, Spain; \\ Instituto de Estructura de la Materia, CSIC, Serrano 121, 28006 Madrid, Spain.
	E-mail: \email{gonzalo.olmo@csic.es}}

\received{\today} 		
\accepted{}		

\preprint{}	
  
\abstract{
We show that an invariant an universal length scale can be consistently introduced in a generally covariant theory through the gravitational sector using the Palatini approach. The resulting theory is able to capture different aspects of quantum gravity phenomenology in a single framework. In particular, it is found that in this theory field excitations propagating with different energy-densities perceive different background metrics, which is a fundamental characteristic of the DSR and Rainbow Gravity approaches. We illustrate these properties with a particular gravitational model and explicitly show how the {\it soccer ball problem} is avoided in this framework. The isotropic and anisotropic cosmologies of this model also avoid the big bang singularity by means of a big bounce. }

\keywords{Palatini Formalism, Quantum Gravity Phenomenology, Non-singular Cosmologies, Cosmic Bounce, Planck Scale}

\begin{document} 

\section{Introduction}

 The consideration of gravitational phenomena in a material world with relativistic quantum properties suggests that Newton's and Planck's constants may be combined with the speed of  light to generate a length $l_P=\sqrt{\hbar G/c^3}\sim 10^{-35}$m, 
 which is known as the Planck length. The Planck length is usually interpreted as the scale at which quantum gravitational phenomena should play a non-negligible role. However, since lengths are not relativistic invariants, the existence of the Planck length raises doubts about the nature of the reference frame in which it should be measured and about the limits of validity of special relativity itself. To explore the potential effects of the existence of the Planck scale in quantum field theory, phenomenological approaches aimed at combining in the same framework the constancy of the speed of light and also an invariant maximum scale of energy, $E_P=\hbar c/l_P$, 
 have been considered in recent years under the name of deformed or doubly special relativity (DSR) \cite{DSR1,DSR2}. A DSR theory can be constructed assuming that the Lorentz group acts non-linearly on the components of the 4-momentum of a particle in such a way that the energy and/or momentum of the particle never become larger than a given scale (of order $\sim E_P$). This can be achieved by defining the physical 4-momentum $P_a=(-E,\vec{p})$ as related to an auxiliary 4-momentum $\Pi_b=(-\epsilon,\vec{\Pi})$ via a non-linear transformation $P_a=(-h(\epsilon,\Pi),g(\epsilon,\Pi)\vec{\Pi}/\Pi)$, where $\Pi=\sqrt{\Sigma_i\Pi_i^2}$ and $g$ and $h$ are given functions, such that the action of the Lorentz group on $\Pi_b$ is linear.  
For particular choices of $g$ and $h$, one may find interesting results such as predicting an energy-dependent speed of photons, which could have observable consequences, 
potentially making the Planck scale experimentally accessible \cite{AmelinoCamelia:2009pg}.\\
\indent The extension of the DSR ideas to the gravitational sector led to what is known as Rainbow Gravity (RG) \cite{Magueijo:2002xx}. Combining the equivalence principle with the basic postulates of DSR, Magueijo and Smolin designed a framework which should capture some of the expected aspects of quantum geometry at high energies and recover General Relativity (GR) at energies well below the Planck scale. They proposed a one-parameter family of equations of the form
\begin{equation}\label{eq:Gmn-SM}
G_{\mu\nu}(E)=8\pi G(E)T_{\mu\nu}(E)+g_{\mu\nu}\Lambda(E)
\end{equation}
with $ds^2=g_{\mu\nu}(E)dx^\mu dx^\nu$, and $E$ being the energy of the test particle used to probe the geometry.   For a FRW spacetime, the metric needed to compute the new field equations can be written as 
\begin{equation}\label{eq:FRW-RG}
ds^2=-\frac{1}{h^2(E/E_P)}dt^2+\frac{a^2(t)}{g^2(E/E_P)}d\vec{x}^{\ 2} \ ,
\end{equation}
and for a spherically symmetric, static spacetime one can use 
\begin{equation}\label{eq:Sch-RG}
ds^2=-\frac{A(r)}{h^2(E/E_P)}dt^2+\frac{1}{g^2(E/E_P)}\left(\frac{dr^2}{B(r)}+r^2 d\Omega^2\right) \ . 
\end{equation}
Equations (\ref{eq:Gmn-SM}) imply that the geometry of spacetime becomes energy dependent, that quanta of different energies see different classical geometries, as explicitly shown in the ansatz (\ref{eq:FRW-RG}) and (\ref{eq:Sch-RG}). In general one assumes that $h$ and $g$ tend to unity when $E/E_P\to 0$ to agree with GR at low energies, and that $h\neq g$ to allow null particles to perceive the energy-dependent effects. \\  
\indent Though the DSR and RG approaches have interesting features that make contact with various aspects of the expected phenomenology of quantum gravity, they also have several disturbing aspects or deficiencies. 
{\bf i)} Their implementation gives a dominant role to the principle of relativity, being gravity and the quantum mere spectators in DSR that appear only through the definition of the Planck energy. {\bf ii)} The field equations of RG cannot be derived from a single action. A one parameter family of such actions is necessary. {\bf iii)} The dependence of the metric components on the energy of the test particle becomes disturbing when the probe is a macroscopic object ($E\gg E_P$).  \\
\indent In this work we present an action-based theory of gravity that exhibits many of the qualitative energy-dependent features of RG and recovers a DSR-like behavior in the appropriate limit but lacks of the problems mentioned above. We provide an explicit model that illustrates all these aspects and which, in addition, has been shown to resolve the big bang singularity in a way closely related to the effective dynamics of loop quantum cosmology \cite{Olmo-Singh09,lqc}.

\section{Motivating the Palatini Approach}

\indent To combine in the same framework the speed of light and the Planck length in a way that preserves the invariant and universal 
nature of both quantities, we first note that though $c^2$ has dimensions of squared velocity it represents a 4-dimensional Lorentz scalar rather than the squared of a privileged 3-velocity. Similarly, we may see $l_P^2$ as a 4-d invariant with dimensions of length squared that, as opposed to the DSR/RG view, needs not be related with any privileged 3-length. Because of dimensional compatibility with a curvature, the invariant $l_P^2$ could be introduced in the theory via the gravitational sector by considering departures from GR at the Planck scale motivated by quantum effects. However, the situation is not as simple as it may seem at first. In fact, an action like
\begin{equation}\label{eq:f(R)}
S[g_{\mu\nu},\psi]=\frac{\hbar}{16\pi l_P^2}\int d^4x \sqrt{-g}\left[R+l_P^2 R^2\right]+S_m[g_{\mu\nu},\psi] \ ,
\end{equation}
where $S_m[g_{\mu\nu},\psi]$ represents the matter sector, contains the scale $l_P^2$ but not in the invariant form that we wished. The reason is that the field equations that follow from (\ref{eq:f(R)}) are equivalent to those of the following scalar-tensor theory
\begin{equation}\label{eq:f(R)-met}
S[g_{\mu\nu},\varphi,\psi]=\frac{\hbar}{16\pi l_P^2}\int d^4x \sqrt{-g}\left[(1+\varphi) R-\frac{1}{4l_P^2}\varphi^2\right]+S_m[g_{\mu\nu},\psi] \ ,
\end{equation}
which given the identification $\phi=1+\varphi$ coincides with the case $w=0$ of Brans-Dicke theory with a non-zero potential $V(\phi)=\frac{(\phi-1)^2}{4l_P^2}$. As is well-known, in this Brans-Dicke theory the observed Newton's constant is promoted to the status of field: $G_{eff}= G/\phi$. The scalar field allows the effective Newton's constant $G_{eff}$ to dynamically change in time and in space according to the equation
\begin{equation}\label{eq:w=0}
3 \Box \phi +\phi V_\phi-2V=\kappa^2T \ , 
\end{equation}
where $\kappa^2=8\pi l_P^2/\hbar$, and $V_\phi\equiv dV/d\phi$. As a result, the corresponding effective Planck length, $\tilde{l}_P^2=l_P^2/\phi$, can also vary in space and time. This is quite different from the assumed constancy and universality of the speed of light in special relativity, which is  implicit in our construction of the total action. In fact, our action has been constructed assuming the Einstein equivalence principle (EEP), 
whose validity guarantees that the observed speed of light is a true constant and universal invariant, not a field\footnote{If the Einstein equivalence principle is true, then all the coupling constants of the standard model are constants, not fields \cite{Will05}. Their values are fixed save a well-known energy-dependent running due to renormalization. } 
like in varying speed of light theories \cite{VSL}. The situation does not improve if we introduce higher curvature invariants in (\ref{eq:f(R)-met}).  We thus see that the introduction of the Planck length in the gravitational sector in the form of a universal constant like the speed of light is not a trivial issue. The introduction of curvature invariants suppressed by powers of $R_P=1/l_P^2$ unavoidably generates new degrees of freedom which turn Newton's constant into a dynamical field. \\

\indent Is it then possible to modify the gravity Lagrangian adding Planck-scale corrected terms without turning Newton's constant into a dynamical field? The answer to this question is in the affirmative. One must first note that metricity and affinity are a priori logically independent concepts \cite{Zanelli}. If we construct the theory {\it à la Palatini}, that is in terms of a connection not a priori constrained to be given by the Christoffel symbols, then the resulting  equations do not necessarily contain new dynamical degrees of freedom (as compared to GR), and the Planck length may remain space-time independent in much the same way as the speed of light and the coupling constants of the standard model, as required by the {EEP}. A natural alternative, therefore, seems to be to consider (\ref{eq:f(R)}) in the Palatini formulation. The field equations that follow from (\ref{eq:f(R)}) when metric and connection are varied independently are \cite{review}
\begin{eqnarray}
f_R R_{\mu\nu}(\Gamma)-\frac{1}{2}f g_{\mu\nu}&=&\kappa^2 T_{\mu\nu} \label{eq:metric}\\
\nabla_\alpha\left(\sqrt{-g}f_R g^{\beta\gamma}\right)&=&0 \ , \label{eq:connection}
\end{eqnarray}
where $f=R+l_P^2R^2$, and $f_R\equiv \partial_R f=1+2l_P^2R$. The connection equation (\ref{eq:connection}) can be easily solved after noticing that the trace of (\ref{eq:metric}) with $g^{\mu\nu}$, 
\begin{equation}\label{eq:T}
R f_R-2f=\kappa^2T \ ,
\end{equation}
 represents an algebraic relation between $R\equiv g^{\mu\nu}R_{\mu\nu}(\Gamma)$ and $T$, which generically implies that $R=R(T)$ and hence $f_R=f_R(R(T))$ [from now on we denote $f_R(T)\equiv f_R(R(T))$]. It should be noted that Palatini $f(R)$ theories also have a Brans-Dicke formulation with parameter $w=-3/2$ and the same  potential as in the metric formulation. In this case, however, the scalar field satisfies the equation \cite{review}
\begin{equation}\label{eq:w=0}
\phi V_\phi-2V=\kappa^2T \ ,
\end{equation}
which is non-dynamiccal and implies that $\phi$ can be expressed as an algebraic function of the trace $T$, $\phi=\phi(T)$. This is just a manifestation of the generalized relation $R=R(T)$ discussed above but using a different notation. Because of this algebraic, non-dynamical, behavior of the scalar field, the theory has no new degrees of freedom and Newton's constant remains a constant, not a field, as we required to keep the speed of light and the Planck length as universal invariants. \\
For the particular Lagrangian (\ref{eq:f(R)}), we find that $R=-\kappa^2T$, like in GR. This relation implies that (\ref{eq:connection}) is just a first order equation for the connection that involves the matter, via the trace $T$, and the metric. The connection turns out to be the Levi-Civita connection of an auxiliary metric $h_{\mu\nu}$, 
\begin{equation}
\Gamma^\alpha_{\mu\nu}=\frac{h^{\alpha\beta}}{2}\left(\partial_\mu h_{\beta\nu}+\partial_\nu h_{\beta\mu}-\partial_\beta h_{\mu\nu}\right) \ ,
\end{equation}   
which is conformally related with the physical metric, $h_{\mu\nu}= f_R(T) g_{\mu\nu}$. Now that the connection has been expressed in terms of $h_{\mu\nu}$, we can rewrite (\ref{eq:metric}) as follows
\begin{equation}\label{eq:Gmn-pal}
G_{\mu\nu}(h)=\frac{\kappa^2}{f_R(T)} T_{\mu\nu}+\Lambda(T)h_{\mu\nu}
\end{equation}   
where $\Lambda(T)\equiv (f-R f_R)/(2f_R^2)=-(\kappa^2 T)^2/R_P$, and we use $R_P=1/l_P^2$ to remark that $\Lambda(T)$ is strongly suppressed by an inverse power of the Planck curvature. This way of writing the Palatini field equations highlights the formal similitudes between them and the RG equations (\ref{eq:Gmn-SM}). The factor $\frac{\kappa^2}{f_R(T)}\equiv 8\pi G/(1-2\kappa^2T/R_P)$ is similar to $8\pi G(E)$, and the function $\Lambda(T)$ could be seen as a $T$-dependent cosmological {\it constant}.
In analogy with RG, the $T-$dependence of  $\frac{\kappa^2}{f_R(T)}$ could be seen as the expectation that the effective gravitational coupling could depend on the energy scale and satisfy a renormalization group equation, though here $T$ is a (diffeomorphism invariant) scalar dependent on the energy-density of the probe rather than on its total energy. \\

\subsection{Spherically symmetric source plus a test particle. \label{sec:SSSS}}

\indent From the structure of the field equations (\ref{eq:Gmn-pal}) and the relation $g_{\mu\nu}=(1/f_R) h_{\mu\nu}$, it follows that $g_{\mu\nu}$ is affected by the matter-energy in two different ways. The first contribution corresponds to the cumulative effects of matter, and the second contribution is due to the dependence on the local density distributions of energy and momentum. This can be seen by noticing that the structure of the equations (\ref{eq:Gmn-pal}) that determine $h_{\mu\nu}$ is similar to that of GR, which implies that $h_{\mu\nu}$ is determined by integrating over all the sources (gravity as a cumulative effect).  Besides that, $g_{\mu\nu}$ is also affected by the local sources via the factor $f_R^{-1}(T)$, which is formally analogous to the energy dependence characteristic of DSR/RG theories.\\
To illustrate this point in a quantitatively precise manner, let us consider a spherically symmetric, non-rotating presureless body such as a rocky planet or a gold sphere, for example. For such objects a solution for the metric  can be easily obtained using the following ansatz
$ds^2=g_{\mu \nu }dx^\mu dx^\nu=g_{tt}dt^2+g_{rr}dr^2+r^2d\Omega^2=\frac{1}{f_R}h_{\mu \nu }d\tilde{x}^\mu d\tilde{x}^\nu$ such that
\begin{equation}\label{eq:metric-int}
ds^2=\frac{1}{f_R(T)}\left[-A(\tilde{r})e^{2\Phi(\tilde{r})}dt^2+\frac{1}{A(\tilde{r})}d\tilde{r}^2+\tilde{r}^2d\Omega ^2\right] \ ,
\end{equation}
where besides the conformal transformation $g_{\mu\nu}=\frac{1}{f_R}h_{\mu \nu }$ we have introduced a new radial coordinate $\tilde{r}$ such that $\tilde{r}^2=r^2f_R$, being $r$ the usual radial coordinate in the physical frame. We then find 
\begin{eqnarray}\label{eq:Phi0}
\frac{2}{\tilde{r}}\frac{d\Phi}{d\tilde{r}}&=&\frac{\kappa^2}{f_R^2}\left(\frac{{T_{{r}}}^{{r}}-{T_t}^t}{A}\right) \\
-\frac{1}{\tilde{r}^2}\frac{d(\tilde{r}[1-A])}{d\tilde{r}}&=& \frac{\kappa^2{T_t}^t}{f_R^2}+\Lambda(T) \label{eq:A}
\end{eqnarray}
Defining now $A(\tilde{r})=1-2G M(\tilde{r})/\tilde{r}$ in (\ref{eq:A}),
we can rewrite $M(\tilde{r})$ and $\Phi(\tilde{r})$ as
\begin{eqnarray}\label{eq:M}
M(\tilde{r})&=&-\frac{{\kappa}^2}{2{G}}\int_0^{\tilde{r}} dx \ x^2 \left[\frac{{T_t}^t}{f_R^2}+\frac{\Lambda(T)}{\kappa^2}\right]\\
\Phi(\tilde{r})&=&\frac{{\kappa}^2
}{2}\int^{\tilde{r}}_0dx \ x \left[\frac{{T_{\tilde{r}}}^{\tilde{r}}-{T_t}^t}{f_R^2 A}\right]
\label{eq:Phi}
\end{eqnarray}
If we consider a point outside of the sources at radius $r$, where $f_R=1-2\kappa^2T|_{T=0}=1$ implies $\tilde{r}=r$, the above equations can be readily integrated leading to
\begin{eqnarray}\label{eq:M-ext}
M(r)&=&M_\odot-\frac{{\Lambda}(0)}{6{G}}r^3\\
\Phi(r)&=&\Phi_0 \label{eq:Phi-ext} \ , 
\end{eqnarray}
where ${\Lambda}(0)=0$ for our quadratic model, and $M_\odot$ and $\Phi_0$ are constants. Since we are assuming a presureless fluid, ${T_t}^t=-\rho(r)$, using the definition ${\kappa}^2\equiv 8\pi {G}$ and our quadratic $f(R)$ model we find that $M_{\odot}$ and $\Phi_0$ are given by
\begin{eqnarray}\label{eq:Modot}
M_\odot &=&\int_0^{R_\odot} d\tilde{r}4\pi \tilde{r}^2\rho\left[ \frac{1}{\left(1+\frac{2\rho}{\rho_P}\right)^2}+\frac{\rho}{\rho_P} \right] \\
\Phi_0 &=& {G}\int_0^{R_\odot} d\tilde{r} \frac{4\pi \tilde{r}^2\rho}{(\tilde{r}-2{G}M(\tilde{r}))}\frac{1}{\left(1+\frac{2\rho}{\rho_P}\right)^2}
\end{eqnarray}
where $R_\odot$ is the radius of the object (where $\rho$ vanishes), and we have defined $\rho_P=R_P/\kappa^2=m_P/l_P^3\sim 10^{91}$g/cm$^3$ as the Planck density. Due to the enormous density scale $\rho_P$, the above expressions for $M_\odot$ and $\Phi_0$ naturally recover the results of GR for regular sources of matter (even for neutron stars, where $\rho_{ns}\sim 10^{14}$g/cm$^3$). Therefore, if outside of the central distribution of matter there are no other sources or test particles, the metric takes the following (exact) form 
\begin{equation}\label{eq:Schw}
ds^2=-\left(1-\frac{2GM_\odot}{r}\right)d{t}^2+\frac{dr^2}{\left(1-\frac{2GM_\odot}{r}\right)}+r^2d\Omega^2 \ ,
\end{equation}
which is the well-known Schwarzschild metric (we have absorbed the constant factor $e^{2\Phi_0}$ in a redefinition of the time coordinate). \\
Consider now the propagation on this geometry of a particle of total energy $E\ll M_\odot$ whose energy density is distributed according to a given function $\rho_{test}$. For a non-relativistic particle or wave-packet described by the  Schr\"{o}dinger equation, for instance, one finds that $T\approx -\rho_{test}=-m_{test} |\psi(t,\vec{x})|^2$, where $|\psi(t,\vec{x})|^2$ represents the probability density, such that $\int d^3x |\psi(t,\vec{x})|^2=1$, and the dominant contribution to the energy comes from the rest mass, $E\approx\int d^3x \rho_{test}=m_{test}$. 
 According to (\ref{eq:M}), the energy of the particle modifies the geometry by its contribution to the total ${T_t}^t$ . However, since we are assuming $E\ll M_\odot$, its contribution to the functions $A(\tilde{r})$ and $\Phi(\tilde{r})$ should be negligible. In other words, its contribution to the $h_{\mu\nu}$ part of the metric is negligible. On the other hand, the physical metric $g_{\mu\nu}$ also depends on the local energy density through the conformal factor $1/f_R(T)$. This means that along the particle trajectory, the factor $\rho/\rho_P$ that appears in $f_R=1+2\rho/\rho_P$ induces on the metric a local dependence on the energy-density of the particle. If we neglect this effect on the functions $A(\tilde{r})$ and $\Phi(\tilde{r})$ (which should have contributions of order $\frac{M_0}{r}\frac{\rho_{test}}{\rho_P}$ due to the coordinate relation $\tilde{r}^2=r^2(1+2\rho_{test}/\rho_P)$), the metric perceived by our test particle is of the form\footnote{Strictly speaking, this line element assumes that our wave-packet represents a spherical wave with support along some trajectory $r=\lambda(t)$ and a certain width $\epsilon$. One could simplify this kind of construction by writing (\ref{eq:Schw}) in Cartesian isotropic coordinates. The wave packet could then be described, for instance, as a spherical distribution of radius $\epsilon$ localized along a trajectory $\vec{x}=\vec{x}(t)$, which is closer to the intuitive view of a point particle propagating through space-time. To avoid unessential technical complications, we keep our discussion within the simple spherically symmetric solution described above.}
\begin{equation}\label{eq:Sch-Pal}
ds^2=-\frac{\left(1-\frac{2GM_\odot}{r}\right)}{\left(1+\frac{2\rho_{test}}{\rho_P}\right)}d{t}^2+\frac{\left(1+\frac{2}{\rho_P}\frac{d(r\rho_{test})}{dr}\right)^2}{\left(1+\frac{2\rho_{test}}{\rho_P}\right)\left(1-\frac{2GM_\odot}{r}\right)}dr^2+r^2d\Omega^2 \ .
\end{equation}
This equation should be compared with (\ref{eq:Sch-RG}) to note the formal similarities between them\footnote{Note that (\ref{eq:Sch-RG}) is expressed in spherical isotropic coordinates, whereas (\ref{eq:Sch-Pal}) appears in Schwarzschild-like coordinates. The apparent dependence of the size of the two-spheres in (\ref{eq:Sch-RG}) on the energy of the test-particle is just an effect of the choice of coordinates.}. The dependence of the line element (\ref{eq:Sch-RG}) on the energy of the probe is now replaced by a dependence on its energy-density. This property is analogous to that described in RG, where particles of different energies (energy-densities in our case) perceive different metrics. In this sense, a clear improvement of our theory is that any macroscopic object with energy above the Planck mass but whose density is well below $\rho_P$ will not perceive any substantial deformation of the geometry as compared to general relativity. This can be extended beyond our particular example to any spacetime containing
 sources of low energy-density as compared to the Planck scale ($|\kappa^2 T/R_P|\ll 1$). For the quadratic model $f(R)=R+R^2/R_P$, in this region (\ref{eq:Gmn-pal}) boils down to $G_{\mu\nu}(h)=\kappa^2T_{\mu\nu}+O(\kappa^2T/R_P)$, and $h_{\mu\nu}\approx (1+O(\kappa^2T/R_P))g_{\mu\nu}$, which implies that the GR solution is a very good approximation. This argument, together with our explicit example, confirm that $h_{\mu\nu}$ is determined essentially by an integration over the sources, like in GR. If this region is traversed by a particle of mass $m\ll M_{Tot}$ but with a non-negligible ratio $\kappa^2T/R_P$, then the contribution of this particle to $h_{\mu\nu}$ can be neglected, but its effect on $g_{\mu\nu}$ via de factor $f_R^{-1}=1-\kappa^2T/R_P$ on the region that supports the particle (its classical trajectory) may be important.\\

Another interesting aspect that can be addressed using the explicit solution presented here is related to the concept of test-particle. In GR test particles are usually seen as structureless, neutral objects which follow geodesics of the metric because their mass/energy is sufficiently small not to significantly modify the background metric. The energy density of such objects is sometimes represented by means of a Dirac delta function along their classical trajectory (even though such configurations would imply the formation of black holes!). Such description is totally inadequate in our approach because the metric has an explicit dependence on the ratio $\rho_{test}/\rho_P$ (or, more generally, on $T/\rho_P$). It is for this reason that we have made explicit the form $\rho_{test}=m_{test}|\psi(t,\vec{x})|^2$, because the form in which the energy density of the particle is distributed matters for the determination of the metric seen by the particle. Particles and field excitations must be generally treated as extended objects with a smooth and finite density profile. If the energy of a given particle (or wave-packet of field excitations) is concentrated on a very small region such that its energy density is close to $\rho_P$, then the metric seen by this particle will be substantially distorted as compared to that seen by a low-density wave-packet. This phenomenon will surely lead to dispersion effects induced by the gravitational backreaction of the space-time domain that supports the particle, which could set limits on the energy density of ultrahigh-energy cosmic rays. A quantitative analysis of these aspects lies beyond to purpose of this work and will be explored elsewhere.\\

\section{Beyond $f(R)$}

\indent The Palatini version of the quadratic $f(R)$ theory considered above captures most of the elements present in RG and naturally cures the so-called {\it soccer ball problem} for macroscopic bodies thanks to the dependence on the local energy-momentum density rather than on the total energy-momentum of the particle (see \cite{sabine} for a related discussion and \cite{Relative-Locality} for a novel approach that also solves the soccer ball problem\footnote{It should be noted that the solution to the {\it soccer ball problem} presented here is technically and conceptually much more economical than that presented in \cite{Relative-Locality}.}). Nonetheless, DSR and RG are characterized by two energy-dependent functions $g$ and $h$ which are a priori independent and, more importantly, allow for energy-dependent photon trajectories. The conformal relation existing between the metrics $h_{\mu\nu}$ and $g_{\mu\nu}$ is characterized by the single function $f_R(T)$ and implies that  particles following null trajectories cannot distinguish between $h_{\mu\nu}$ and $g_{\mu\nu}$. Therefore, to fully capture the phenomenology of RG, the theory should be able to generate density-dependent effects which affect space and time directions differently. Theories with this additional property can also be found within the Palatini approach by just enlarging the class of Lagrangians from $f(R)$ to $f(R,Q)$, where $Q\equiv R_{\mu\nu}R^{\mu\nu}$, and $R_{\mu\nu}$ is the symmetric Ricci tensor of $\Gamma^\lambda_{\alpha\beta}$ \footnote{To avoid the introduction of dynamical degrees of freedom, we restrict the action to depend only on the symmetric part of the Ricci tensor \cite{OSAT09}.  An extra dependence on its antisymmetric part amounts to introducing a massive Proca field \cite{Vitagliano:2010pq}, but it has no effect on the energy-density dependence of the metric. }.  In this case, the field equations for metric and connection are \cite{review}
\begin{eqnarray}
f_R R_{\mu\nu}-\frac{1}{2}f g_{\mu\nu}+2f_Q R_{\mu\alpha}{R^{\alpha}}_{\nu}&=&\kappa^2 T_{\mu\nu} \label{eq:f(R,Q)-metric}\\
\nabla_\alpha\left[\sqrt{-g}\left(f_R g^{\beta\gamma}+2f_Q R^{\beta\gamma}\right)\right]&=&0 \ . \label{eq:f(R,Q)-connection}
\end{eqnarray}
where $f_Q\equiv \partial_Q f$. 
Defining the tensor ${P_\mu}^\nu=R_{\mu\alpha}g^{\alpha\nu}$, (\ref{eq:f(R,Q)-metric}) can be seen as a matrix equation, 
\begin{equation}\label{eq:matrix}
2f_Q {P_\mu}^\alpha{P_\alpha}^\nu+f_R{P_\mu}^\nu-\frac{1}{2}f {\delta_\mu}^\nu=\kappa^2 {T_\mu}^\nu \ ,
\end{equation}
which establishes an algebraic relation between the components of ${P_\mu}^\nu$ and those of ${T_\mu}^\nu=T_{\mu\alpha}g^{\alpha\nu}$, i.e., ${P_\mu}^\nu={P_\mu}^\nu({T_\alpha}^\beta)$. Once the solution of (\ref{eq:matrix}) is known, one can express $R$ and $Q$ in terms of the matter according to the identities $R=Tr[P]={P_\mu}^\mu$ and $Q=Tr[P^2]={P_\mu}^\alpha{P_\alpha}^\mu$. Solutions of (\ref{eq:matrix}) have been found for quadratic models of the form $f(R,Q)=R+aR^2/R_P+Q/R_P$ with ${T_\mu}^\nu$ represented by a perfect fluid \cite{OSAT09,BO2010}. \\
The relation ${P_\mu}^\nu={P_\mu}^\nu({T_\alpha}^\beta)$ implies that $R$ and $Q$ are functions of ${T_\mu}^\nu$ (not just of the trace $T$), which allows to proceed similarly as in the $f(R)$ case and solve for the independent connection as the Levi-Civita connection of a new auxiliary metric $\tilde{h}_{\mu\nu}$ (see \cite{OSAT09} for details) which is related to $g_{\mu\nu}$ by means of the following non-conformal relation
\begin{equation}\label{eq:hmn-general}
\tilde{h}^{\mu\nu}=\frac{g^{\mu\alpha}{\Sigma_\alpha}^\nu}{\sqrt{\det \Sigma}} \ ,
\end{equation}
where 
\begin{equation}
{\Sigma_\alpha}^\nu=f_R\delta_\alpha^\nu+2f_Q {P_\alpha}^\nu \ 
\end{equation}
is a function of ${T_\mu}^\nu$ and, therefore, depends on the local densities of energy and momentum.  The relation (\ref{eq:hmn-general})
allows to write the field equation (\ref{eq:f(R,Q)-metric}) in the following compact form
\begin{equation}\label{eq:Rmn-f(R,Q)}
{R_\mu}^\nu(\tilde{h})=\frac{1}{\sqrt{\det\hat\Sigma}}(\frac{f}{2}{\delta_\mu}^\nu +\kappa^2 {T_\mu}^\nu) \ ,
\end{equation}
where ${R_\mu}^\nu(\tilde{h})\equiv R_{\mu\alpha}(\tilde{h}) \tilde{h}^{\alpha\nu}$ and ${T_\mu}^\nu\equiv T_{\mu\alpha} g^{\alpha\nu}$. The dependence of this equation on $Q$, not just on $R$, guarantees that there is modified dynamics even for sources with $T=0$, such as the electromagnetic field and a gas of radiation. It should be noted that this equation is equivalent to (\ref{eq:Gmn-pal}) in the $f(R)$ limit, i.e., when $f_Q=0$. \\

It is important to note that (\ref{eq:f(R,Q)-metric})  (or, equivalently, (\ref{eq:Rmn-f(R,Q)})) in vacuum (${T_\mu}^\nu=0$) boils down exactly to the equations of GR with (possibly) an effective cosmological constant (depending on the form of the Lagrangian). This can be seen by rewriting (\ref{eq:matrix}) in vacuum as 
\begin{equation}
2f_Q\left(\hat{P}+\frac{f_R}{4f_Q}\hat{I}\right)^2=\left(\frac{f^2_R}{8f_Q}+\frac{f}{2}\right)\hat{I}  \ ,
\end{equation}
where $\hat{P}$ and $\hat{I}$ denote the matrices ${P_\mu}^\nu$ and ${\delta_\mu}^\nu$, respectively. The physical solution  to this equation, which recovers the $f(R)$ theory in the limit $f_Q\to 0$, is of the form 
\begin{equation}
{P_\mu}^\nu=-\frac{f_R}{4f_Q}\left(1-\sqrt{1+\frac{4f_Q f}{f_R^2}}\right){\delta_\mu}^\nu\equiv \Lambda(R,Q){\delta_\mu}^\nu  \ .
\end{equation}
This equation can be used to compute $R_0\equiv{P_\mu}^\mu|_{vac}= 4\Lambda(R_0,Q_0)$ and $Q_0={P_\mu}^\alpha {P_\alpha}^\mu|_{vac}=4\Lambda(R_0,Q_0)^2$, which lead to the characteristic relation $Q_0=R^2_0/4$ of de Sitter spacetime. 
For the quadratic models $f(R,Q)=R+aR^2/R_P+Q/R_P$, for instance, one can also use the trace of (\ref{eq:Rmn-f(R,Q)}) to find that $R_0=0$, from which $Q_0=R_0^2/4=0$ follows. For a generic $f(R,Q)$ model, in vacuum  one finds that ${\Sigma_\mu}^\nu= a(R_0){\delta_\mu}^\nu$ and $\tilde{h}_{\mu\nu}= a(R_0) g_{\mu\nu}$, with $a(R_0)=f_R\left(1+\sqrt{1+\frac{4f_Q f}{f_R^2}}\right)/2$ evaluated at $R_0$. Therefore, in vacuum (\ref{eq:Rmn-f(R,Q)}) can be written as ${R_\mu}^\nu(\tilde{h})={R_\mu}^\nu(g)=\Lambda_{eff}{\delta_\mu}^\nu$, with $\Lambda_{eff}=f(R_0,Q_0)/2 a(R_0)^2$, which shows that the field equations coincide with those of GR with an effective cosmological constant.\\

\subsection{Spherically symmetric source in $f(R,Q)$ theories. \label{sec:SSSS2}}

The above considerations about the empty space dynamics are necessary to understand that the family of quadratic $f(R,Q)$ Lagrangians admits the solution (\ref{eq:Schw}) outside of the sources. Obviously, the integral representation for the constants $M_\odot$ and $\Psi_0$ must be adapted to the new theory. This can be done as follows. For a perfect fluid, $T_{\mu\nu}=(\rho+P)u_\mu u_\nu+P g_{\mu\nu}$, the metric $g_{\mu\nu}$ within the fluid can be expressed as \cite{BO2010}
\begin{equation}\label{eq:gmn-f(R,Q)}
g_{\mu\nu}=\frac{1}{\Omega}\tilde{h}_{\mu\nu}+\frac{\Lambda_2}{\Lambda_1-\Lambda_2} u_\mu u_\nu \ ,
\end{equation}
where $\Omega=\left[\Lambda_1(\Lambda_1-\Lambda_2)\right]^{1/2}$, $\Lambda_1= \sqrt{2f_Q}\lambda+\frac{f_R}{2}$, $\Lambda_2= \sqrt{2f_Q}\left[\lambda\pm\sqrt{\lambda^2-\kappa^2(\rho+P)}\right]$, and $\lambda=\sqrt{\kappa^2 P+\frac{f}{2}+\frac{f_R^2}{8f_Q}}$. For the family of models $f(R,Q)=\tilde{f}(R)+Q/R_P$ one finds that $Q$ is given by
\begin{equation}\label{eq:Q}
\frac{Q}{2R_P}=-\left(\kappa^2P+\frac{\tilde f}{2}+\frac{R_P}{8}\tilde f_R^2\right)+\frac{R_P}{32}\left[3\left(\frac{ R}{R_P}+\tilde f_R\right)-\sqrt{\left(\frac{ R}{R_P}+\tilde f_R\right)^2-\frac{ 4 \kappa^2(\rho+P)}{R_P} }\right]^2 \ ,
\end{equation} 
where $R\tilde{f}_R-2\tilde{f}=\kappa^2T$ determines the particular relation $R=R(T)$ for each $\tilde{f}(R)$ model. In particular, for $f(R)=R+a R^2/R_P$, one finds the same relation as in GR, $R=-\kappa^2 T$.\\
The $\tilde{h}_{\mu\nu}$ part of the metric (\ref{eq:gmn-f(R,Q)}) can be obtained using (\ref{eq:Rmn-f(R,Q)}) and integrating over the fluid. 
Defining an auxiliary line element in Schwarzschild-like coordinates for $\tilde{h}_{\mu\nu}$ as 
\begin{equation}\label{eq:ds2tilde}
d\tilde{s}^2=\tilde{h}_{\mu\nu} d\tilde{x}^\mu d\tilde{x}^\nu=-B(\tilde{r})e^{2\Psi(\tilde{r})}dt^2+\frac{1}{B(\tilde{r})}d\tilde{r}^2+\tilde{r}^2d\Omega ^2 \ ,
\end{equation}
with $B(\tilde{r})=1-2G M(\tilde{r})/\tilde{r}$, for a presureless fluid we find ($\sqrt{\det\hat{\Sigma}}=\Omega \Lambda_1$)
\begin{eqnarray}
\label{eq:Psi0}
\frac{d\Psi}{d\tilde{r}}&=&\frac{1}{2\Omega \Lambda_1 }\frac{\kappa^2\rho \tilde{r}}{B} \\
M_{\tilde{r}}&=& \frac{(f+\kappa^2\rho)\tilde{r}^2}{4G\Omega \Lambda_1} \label{eq:B} \ ,
\end{eqnarray}
which are the generalization of (\ref{eq:Phi0}) and (\ref{eq:A}) to the $f(R,Q)$ case. The physical line element $ds^2=g_{tt}dt^2+g_{rr}dr^2+r^2d\Omega^2$ is non-trivially related with the line element $d\tilde{s}^2$ defined in (\ref{eq:ds2tilde}) because the relation between $g_{\mu\nu}$ and $\tilde{h}_{\mu\nu}$ is not conformal. With a bit of algebra, one finds that $g_{tt}=\frac{\tilde{h}_{tt}}{\Lambda_1}\sqrt{\frac{\Lambda_1-\Lambda_2}{\Lambda_1}}$, $g_{rr}=\frac{\tilde{h}_{\tilde{r}\tilde{r}}}{\Omega}\left(\frac{d\tilde{r}}{dr}\right)^2$, and $\tilde{r}^2=r^2\Omega$. These expressions coincide with those of $f(R)$ theories in the limit $f_Q\to 0$, which leads to $\Omega\to f_R$, $\Lambda_1\to f_R$, and $\Lambda_2\to 0$. \\

In parallel with the $f(R)$ case, the lesson to be extracted from this analysis is that $\tilde{h}_{\mu\nu}$ contains the cumulative effects of matter, i.e., it is essentially determined by integrating over the sources, while the $\Lambda_1$,$\Lambda_2$ and $\Omega$ terms that appear in the components of the physical metric provide the local dependence on the matter-energy densities. It is important to note that $\Lambda_1$,$\Lambda_2$ and $\Omega$, which are functions of $R$ and $Q$, are not constant for electromagnetic waves, which contrasts with the $f(R)$ case. For the electromagnetic field, $T=0$ implies that $R=0$ but (\ref{eq:Q}) gives
 $Q=(3R_P^2/8)\left[1-\frac{8\rho_{em}}{3\rho_P}-\sqrt{1-\frac{16\rho_{em}}{3\rho_P}}\right]$. If we consider a spherical wave of radiation with total energy $E\ll M_\odot$ collapsing towards the central source, along its trajectory the metric components $g_{tt}$ and $g_{rr}$ will deviate from the Schwarzschild form by the corresponding density-dependent terms. If the energy density is small, $\rho_{em}/\rho_P\ll 1$ then $Q\approx (4/3)(\kappa^2\rho_{em})^2(1+8\rho/3\rho_P+\ldots)$ does not substantially change the metric. However, there must be important modifications as $\rho_{em}$ approaches its maximum value $\rho_{max}=3\rho_P/16$. 
 One thus expects new dispersion effects for very intense light beams (with energy densities approaching $\rho_P$). The possibility of birefringence effects, which could be experimentally accessible currently and in the near future analyzing light emissions from distant astrophysical sources \cite{Jacobson:2005bg}, in this type of theories is another aspect that will be explored in future works.

\subsection{Scalar particles and DSR limit}
Let us now consider the propagation of a scalar particle in an $f(R,Q)$ Palatini background. 
For a scalar field with kinetic energy $\chi\equiv g^{\mu\nu}\partial_\mu\phi \partial_\nu\phi$ and Lagrangian $\LL=\chi+2V(\phi)$, 
$\tilde{h}_{\mu\nu}$ and $g_{\mu\nu}$ turn out to be related by
\begin{equation}
g_{\mu\nu}=\frac{1}{\Omega} \tilde{h}_{\mu\nu}\ +\frac{\Lambda_2}{\Lambda_1+\chi\Lambda_2} \partial_\mu\phi \partial_\nu\phi  \label{eq:hdown}
\end{equation}
where $\Omega=\left[\Lambda_1(\Lambda_1+\chi\Lambda_2)\right]^{1/2}$, $\Lambda_1=\sqrt{2f_Q}\lambda+\frac{f_R}{2}$, $\Lambda_2={\sqrt{2f_Q}(-\lambda\pm\sqrt{\lambda^2+\kappa^2 \chi})}/{\chi}$, and $\lambda^2=f/2+f_R^2/8f_Q-\kappa^2\LL/2$. 
Given this relation, in regions that support only the scalar field the metric field equations (\ref{eq:f(R,Q)-metric})  can be rewritten as 
\begin{equation}\label{eq:Tmn-sfh}
R_{\mu\nu}(\tilde{h})=\frac{1}{\Lambda_1}\left[\frac{\left(f-\kappa^2\LL\right)}{2\Omega}\tilde{h}_{\mu\nu}+\frac{\Lambda_1\kappa^2}{\Lambda_1+\chi\Lambda_2}\partial_\mu\phi\partial_\nu\phi\right] \ ,
\end{equation}
which also exhibit a structure similar to that of GR. 
For the particular (quadratic) model 
\begin{equation}\label{eq:quadratic}
f(R,Q)=R-\frac{R^2}{2R_P}+\frac{Q}{R_P} \ ,
\end{equation}
 the low energy-density limit $|\kappa^2\LL/R_P|\ll 1$ leads to
\begin{equation}\label{eq:LowLimit}
R_{\mu\nu}(\tilde{h}) \approx \kappa^2\left(\partial_\mu\phi\partial_\nu\phi+{V}\tilde{h}_{\mu\nu}\right) + \frac{\kappa ^4}{R_P}\left[({2V -\tilde{\chi}})\partial_\mu\phi\partial_\nu\phi+\left(\frac{8 V^2+\tilde{\chi}^2}{4}\right)\tilde{h}_{\mu\nu}\right]+ \ldots
\end{equation}
which is in agreement with GR up to corrections of order $O(1/R_P)$. Note that to this order $\tilde{\chi}\equiv \tilde{h}^{\mu\nu}\partial_\mu\phi \partial_\nu\phi\approx {\chi}$. Similarly as in the $f(R)$ case, this indicates that $\tilde{h}_{\mu\nu}$ is mainly determined by integrating over the sources (cumulative effects of gravity), whereas $\Omega$ and the last term of (\ref{eq:hdown}) represent the local energy-density contributions to the metric. These two local contributions provide (more than) enough freedom to generate the effects associated to the functions $h$ and $g$ of DSR and RG. \\
Consider now a scalar particle propagating through empty spacetime. If the total energy carried by the particle is small and such that $\tilde{h}_{\mu\nu}\approx \eta_{\mu\nu}$,  we find that the spacetime metric for the model (\ref{eq:quadratic}) takes the form
\begin{equation}
g_{\mu\nu}\approx \eta_{\mu\nu}+\frac{2}{\rho_P}\left(V\eta_{\mu\nu}+ \partial_\mu\phi \partial_\nu\phi\right) +O\left(\frac{1}{\rho_P^2}\right). \label{eq:hdown-Mink} 
\end{equation}
This metric only keeps the local energy-momentum density dependence induced by the gravitational sector of the theory
and should be understood as the way to obtain the DSR limit from the full gravitational theory. Obviously, this limit may not always be strictly justified because the Newtonian potential corrections to $\tilde{h}_{\mu\nu}$, though tiny, could be of the same order as the density-dependent terms, depending on the field distribution. However, it serves to illustrate that neglecting the cumulative effects of gravity one recovers a DSR-like behavior. \\
\indent From (\ref{eq:hdown-Mink}) we see that the leading order corrections to the Minkowski metric are strongly suppressed by inverse powers of the Planck density, $\rho_P\equiv R_P/\kappa^2$, which indicates that a perturbative study of such contributions in field theories should be feasible at low energy densities. This should provide an idea of the kind of corrections induced by the Planck-scale modified Palatini dynamics on Minkowskian field theories.  In fact, one can use the metric (\ref{eq:hdown-Mink}) to estimate the first-order modifications of the scalar field equation $\Box \phi -V_\phi=0$ due to the local energy-density dependence of the metric. After some lengthy algebra, one finds
\begin{equation}\label{eq:phi-h}
\partial^2\phi-V_\phi\approx 0+\frac{1}{\rho_P}\left[V_\phi\left(2V -3\partial^\alpha \phi  \partial _{\alpha }\phi\right)+2(\partial^\mu\phi\partial^\nu\phi)\partial_\mu\partial_\nu\phi\right] \ ,
\end{equation}
where $\partial^2\equiv \eta^{\mu\nu}\partial_\mu \partial_\nu$. For a massive scalar with $V(\phi)=m^2\phi^2/2$, the term $V_\phi V$ on the right hand side produces the same effect as a $\lambda \phi^4$ interaction in the Lagrangian with $\lambda\equiv m^4/4\rho_P$. The terms involving derivatives of the field are expected to modify the dispersion relation $E^2=m^2+k^2$ when the scalar amplitude is sufficiently high. This contrasts with other approaches to quantum gravity phenomenology where the proposed modifications of the dispersion relations introduce higher powers of $k^2$ but are independent of the field amplitude. The nonlinear dependence on the field amplitude found here is a distinctive characteristic of Palatini theories, which signals the energy-density dependence of its modified dynamics. 

\subsection{Non-singular cosmologies}

From the analysis of the previous section it is unclear if a perturbative treatment will be able to provide substantial new physics to explore and characterize the Palatini modified dynamics through particle physics experiments. A glance at the dynamics of cosmological models may help gain some insight into this point. In this context one observes that the dynamics in quadratic Palatini models is essentially the same as in GR at all times except very near the big bang singularity. In this region, GR predicts an unbounded growth of the matter-energy density, which causes the singularity. In the Palatini models, however, non-perturbative effects arise and cure the singularity by means of a cosmic bounce. To see this, consider the scalar $Q$ in the model (\ref{eq:quadratic}) with a perfect fluid of density $\rho$ and pressure $P$
\begin{equation}\label{eq:Q-1/2}
Q=\frac{3R_P^2}{8}\left[1-\frac{2\kappa^2(\rho+P)}{R_P}+\frac{2\kappa^4(\rho-3P)^2}{3R_P^2}-\sqrt{1-\frac{4\kappa^2(\rho+P)}{R_P}}\right] \ .
\end{equation}
\FIGURE{\epsfig{file=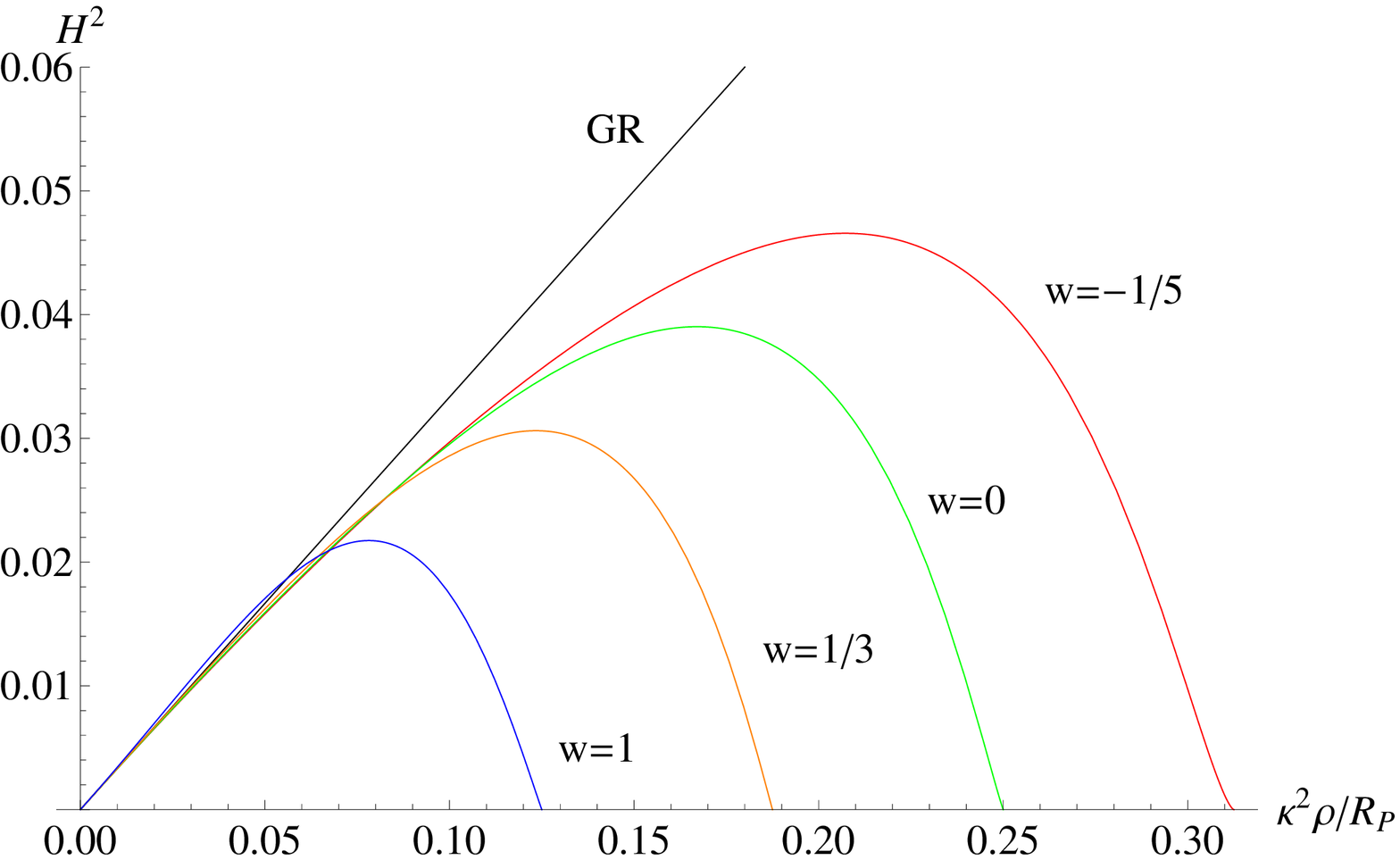, width=0.6\textwidth}
\caption{Hubble function squared in terms of $\rho$ for GR (straight line) and the model $f(R,Q)=R-R^2/2R_P+Q/R_P$. A cosmic bounce occurs at $\kappa^2\rho=R_P/(4+4w)$, where $H^2=0$. \label{fig:am1b2}}
}
At low densities the GR solution is recovered, $Q\approx Q_{GR}+\frac{3 (P+\rho )^3}{2 R_P}$, but as 
$(\rho+P)$ approaches $R_P/4\kappa^2$, important 
departures from GR are expected because negative values of the 
squared root in (\ref{eq:Q-1/2}) are not physically accessible. 
In the case of an isotropic FRW universe with 
constant equation of state $w=P/\rho$, the Hubble function in 
the spatially flat case takes the form \cite{BO2010}
\begin{equation}\label{eq:Hubble-iso}
H^2=\frac{1}{6(\Lambda_1-\Lambda_2)}\frac{\left[f+\kappa^2(\rho+3P)\right]}{\left[1+\frac{3}{2}\Delta_1\right]^2} \ , 
\end{equation}
where $\Delta_1=-(1+w)\rho\partial_\rho \Omega$. At low densities, this expression exactly recovers the linear $\rho$-dependence of GR, 
but near the Planck scale, when $\kappa^2\rho/R_P\sim 0.1$, its behavior changes reaching a maximum and then vanishing at a higher density, which implies a cosmic bounce (see Fig.1). These isotropic bouncing solutions occur for all equations of state comprised within the interval $-1<w\lesssim 11$ and persist in anisotropic (Bianchi I) spacetimes\footnote{It is not known whether this model can avoid other types of singularities or not.} (see \cite{BO2010} for details). It should be noted that the effective dynamics of isotropic loop quantum cosmology \cite{lqc} for a massless scalar was exactly reproduced by an $f(R)$ Palatini theory in \cite{Olmo-Singh09}. The possibility of extending that result to more general spacetimes 
and matter sources using $f(R,Q)$ or other extensions within the Palatini approach is a matter that deserves further study.\\

\section{Summary and Conclusions}

\indent In this work we have shown that an invariant an universal length scale can be consistently introduced in a generally covariant theory through the gravitational sector. This can be done by introducing Planck scale corrections in the gravitational Lagrangian and assuming that metric and connection are independent geometrical entities. The theory so constructed neither leads to higher-order equations nor introduces new dynamical degrees of freedom (such as scalars, vectors, or tensors of different ranks). Rather, it boils down to GR with an effective cosmological constant in vacuum but generates modified dynamics in regions that contain matter/energy sources. In this sense, we have shown that without imposing any a priori phenomenological structure, a quadratic gravitational Lagrangian à la Palatini predicts an energy-density dependence of the metric components that closely matches the structure conjectured by Magueijo and Smolin in \cite{Magueijo:2002xx,DSR2}. The dependence of the metric on the local energy-momentum densities naturally solves the so called {\it soccer ball problem} of DSR and RG. This follows from the fact that a probe whose total energy may be well above the Planck mass may have its energy distributed in such a way that its density never reaches the Planck density scale $\rho_P\sim 10^{91}$ g/cm$^3$. For objects with $\rho/\rho_P\ll 1$, the geometry is essentially the same as in GR. However, objects for which $\rho/\rho_P$ is not small may perceive a metric substantially different from that seen by low density particles. \\

To illustrate in a quantitative manner that the physical metric, $g_{\mu\nu}$, in Palatini theories is made of two different kinds of contributions, we have analyzed spherically symmetric, static distributions of matter. This has allowed us to show that the auxiliary metric $h_{\mu\nu}$ (and $\tilde{h}_{\mu\nu}$) is essentially determined by an integration over the sources, giving account in this way of the cumulative effects of gravity. The other contribution is due to the local energy-momentum densities, and is represented by the conformal factor $1/f_R(T)$ in the case of $f(R)$ theories and by the matrix ${\Sigma_\mu}^\nu ({T_\alpha}^\beta)$ in $f(R,Q)$ theories. On the other hand, we have studied the form of the metric in regions containing a scalar field. We have argued that the DSR limit of the theory can be obtained by neglecting the cumulative effects of matter, i.e., assuming $\tilde{h}_{\mu\nu}=\eta_{\mu\nu}$. We have also used the resulting metric to compute the first order corrections to the scalar field equation due to the modified gravitational dynamics. These corrections can be seen as the gravitational backreaction caused on the field due to its own contribution to the local energy-momentum density. For a massive scalar, they consist on a $\lambda \phi^4$ self-interaction plus modifications of the dispersion relation that depend non-linearly on the amplitude of the field. On the other hand, the analysis of the evolution of cosmological models suggests that the perturbative treatment of these theories could provide no new insights on the properties of Planck scale dynamics. We have seen that the cosmic evolution is almost coincident with that of GR at all times except very near the singularity, where non-perturbative mechanisms act to produce a bounce. \\
In summary, our results suggest that Palatini theories represent a new and powerful framework to address different aspects of quantum gravity phenomenology, which motivates further studies that are currently underway.\\

This work has been supported by the Spanish grants FIS2008-06078-C03-02, FIS2008-06078-C03-03, and the Program CPAN (CSD2007-00042). The author thanks G. Mena-Marugán for very useful and critical discussions during the first stages of the elaboration of this work.


\end{document}